\documentclass[fleqn,usenatbib]{mnras}

\usepackage{newtxtext,newtxmath}

\usepackage[T1]{fontenc}
\usepackage{ae,aecompl}

\usepackage{graphicx}	
\usepackage{amsmath}	
\usepackage{rotating}
\usepackage{listings}
\usepackage{threeparttable}

\title[Soft $\gamma$-ray selected giant  galaxies: an update]{Soft $\gamma$-ray selected giant radio galaxies: an update}

\author[L. Bassani et al.]
{
L. Bassani,$^{1}$\thanks{Contact e-mail:\href{mailto:loredana.bassani@inaf.it}
{loredana.bassani@inaf.it}}
F. Ursini$^{2}$, 
A. Malizia$^{1}$,
G. Bruni$^{3}$,
F. Panessa$^{3}$, 
N. Masetti$^{1,5}$,
I. Saviane$^{4}$,
\newauthor
L. Monaco$^{5}$,
T. Venturi$^{6}$, 
D. Dallacasa$^{7,6}$, 
A. Bazzano$^{3}$, 
and P. Ubertini$^{3}$  \\
$^{1}$ INAF/OAS Bologna, Via P. Gobetti 101, I-40129 Bologna, Italy \\
$^{2}$ Dipartimento di Matematica e Fisica, Universit\'a Roma Tre, via della Vasca Navale 84, I-00146 Roma, Italy\\
$^{3}$ INAF/IAPS Roma, Via Fosso del Cavaliere 100, I00133 Roma, Italy\\
$^{4}$ European Southern Observatory, Alonso de Cordova, 3107, Santiago, Chile\\
$^{5}$ Departamento de Ciencias F\'isicas, Universidad Andr\'es Bello, Fern\'andez Concha 700, Las Condes, Santiago, Chile.\\
$^{6}$ INAF/IRA Bologna,  Via P. Gobetti 101, I-40129 Bologna, Italy \\
$^{7}$Dipartimento di Fisica e Astronomia, Universit\'a di Bologna, Via Gobetti 93/2, 40129 Bologna, Italy\\
}

\date{Accepted XXX. Received YYY; in original form ZZZ}

\pubyear{}

\newcommand{\Nii}{[N {\sc ii}]}

\begin{document}
\label{firstpage}
\maketitle

\begin{abstract}
We present an update on the sample of soft gamma-ray selected giant radio galaxies (GRGs) extracted from INTEGRAL/IBIS
and Swift/BAT surveys; it includes 8 new sources and one candidate object. In the new sample all, but one source, display  FR II radio morphologies; the only exception is B21144+35B which is an FR I. 
The objects belong to both type 1 and 2 AGN optical classes and have  redshifts 
in the range 0.06-0.35,  while  the radio sizes span from 0.7 to 1 Mpc.
In this study, we present for the first time two objects that were never discussed as GRGs before and propose a new candidate GRG.
We confirm the  correlation between the X-ray luminosity and the radio core luminosity found for other soft gamma-ray  selected GRGs 
and expected for AGNs powered by efficient accretion. We also corroborate previous results that indicate that the luminosity of the radio lobes is relatively  low compared with the nuclear X-ray emission. This supports the idea  that  the nucleus of these GRGs is now more powerful than in the past, consistent with a restarting activity scenario.
\end{abstract}


\section{Introduction}
A small fraction of radio galaxies  exhibits exceptional linear extents, i.e. above 0.7 Mpc (for $H_0$= 71 km s$^{-1}$ Mpc$^{-1}$). 
Defined as Giant Radio Galaxies (GRGs), these objects represent the largest and most energetic single entities in the Universe and are of particular interest as extreme examples of radio source development and evolution. Detailed studies of the age of some GRGs suggest that their lobes formed  during tens, possibly hundreds, of Myr (see e.g. \citealt{2020MNRAS.494..902B} and references therein); with such long period of activity  they are the ideal targets to study the duration of the  radio phase in AGN and its duty cycle. 
GRGs are also interesting for many aspects of astrophysical research: as unique laboratories where to study particle acceleration processes and understand cosmic magnetism \citep{2004ApJ...604L..77K,2020arXiv200405169S}, to study the formation of large-scale structures  and also to probe   the Warm-Hot Intergalactic Medium interacting with the magnetized relativistic plasma \citep{2013MNRAS.432..200M}. Recently the study of GRGs has gained momentum thanks to LOFAR observations, which allow to discover  new objects \citep{2020A&A...635A...5D} providing at the same time  interesting insights into  their nature \citep{2020arXiv200503708D}.

Starting from 2002, the soft gamma-ray sky has been surveyed by INTEGRAL/IBIS and subsequently by Swift/BAT at energies above  10 keV. Up to now various all sky catalogues have been released  revealing a large population of active galactic nuclei. More specifically, around 6-8\% of these soft gamma-ray selected AGNs display a double lobe morphology typical 
of  radio galaxies \citep{2016MNRAS.461.3165B}. What is intriguing and particularly interesting is that a substantial fraction (25$\%$) of these radio galaxies have giant radio structures: considering that
the overall fraction of giants in samples of radio galaxies is much lower 
 ($\sim$ 6$\%$ in the 3CR catalogue, \citealt{1999MNRAS.309..100I} and $\sim$ 8$\%$ among LOFAR objects, \citealt{2020A&A...635A...5D}),
the fraction found in soft gamma-ray surveys is impressive, and suggests a tight
connection between the nuclear/accretion properties of the AGN and the radio source size.
At the moment 14 GRGs\footnote{See GRACE at https://sites.google.com/inaf.it/grace/home, for information on this first sample} have been already found   within INTEGRAL/IBIS and Swift/BAT surveys  \citep{2016MNRAS.461.3165B}, but these databases are constantly being updated allowing the identification of new soft gamma-ray selected objects.
After their initial finding, \citet{2016MNRAS.461.3165B} put a large observational effort to study the first
soft gamma-ray selected sample of 14 GRGs,  finding evidence of restarting radio activity in almost every
object given that some GRGs host a young radio core (Giga Hertz/High Frequency Peaked  or GPS/HFP,
\citealt{2019ApJ...875...88B}) while   others display a double-double and/or X-shaped morphology
\citep{2020MNRAS.494..902B}.
Note that the radio-emitting electrons in GPS/HFP sources usually have a radiative age of a few kyr, while the Mpc-scale structure of GRGs usually takes tens/hundreds of Myr to develop. Furthermore,  a double-double  morphology implies a source showing two  pairs of lobes around a common core while  an X shaped one could be the result  of a drastic jet axis flip  producing  new lobes with different orientation on top of relic ones (but see the case of NCG326, \citealt{2019MNRAS.488.3416H}): overall both morphologies can be taken as signs of  multiple episodes of activity.
Our team also found that one GRGs  underwent a jet axis change of about  90 degrees and is now pointing towards us like a blazar \citep{2017A&A...603A.131H}; another source was found to be  enclosed  in a diffuse radio emitting cocoon, relic of a previous active phase \citep{2020MNRAS.494..902B}. Overall these results imply that soft gamma-ray selected GRGs are extremely interesting objects and important laboratories where to study AGN duty cycle.

In this paper, we report an update to the original list of soft gamma-ray selected GRGs; this new sample lists 8 more sources plus a candidate one, including 2 (possibly 3 if the candidate is confirmed) new giant radio galaxies. For each radio galaxy we  measured the largest angular size (LAS) in arcsec and then calculated the corresponding
projected linear size in Mpc at the source's redshift assuming  standard cosmological parameters ($H_0$= 71 km s$^{-1}$
Mpc$^{-1}$, $\Omega_{\rm }$=0.27, $\Omega_{\Lambda}$=0.73) consistent  with our previous work \citep{2016MNRAS.461.3165B}.

The radio as well as the high energy properties of the sample are also presented and discussed.

\begin{table*}
\footnotesize
\begin{threeparttable}[b]
\caption{New  soft gamma-ray selected Giant Radio Galaxies.The columns from 1 to 10  report the source name (soft gamma-ray plus another from SIMBAD), redshift,
optical class,  radio morphology, 14-195 keV soft gamma-ray luminosity, conversion factor from arcsec to kpc,  Largest Angular Size(LAS), 
previous  references to the giant nature and radio image (if available)  and  finally the radio size (in Mpc).}
\centering
\begin{tabular}{l l c c c c c c c c }
\hline\hline
Soft $\gamma$-ray Name  &   Other Name & z        &  Opt Class  & Radio Morph  & Log L $_{\textrm{14-195 keV}}$   & Conv Fac       & LAS      & Ref    & Size  \\
                       &               &          &              &             & erg/s                            &  kpc/arcsec    & arcsec   &        &   Mpc   \\
\hline\hline
SWIFT J0632.1-5404   & QSO J0631-5404&  0.2036 &    Sey1     & FR II       & 44.90                           &   3.314        & 310      & 1,a   &  1.03  \\ 
SWIFT J0636.5-2036  & PKS 0634-20   &  0.0552 &    Sy2      & FR II       & 43.80                           &   1.059        & 900      & 1,b   &  0.95  \\
SWIFT J0801.7+4764   & RBS 0688      &  0.1567 &    Sy1     & FR II       & 44.70                           &  2.682        & 360      & 1,c   &  0.97  \\
B2 1144+35B          & RGB J1147+350 &  0.0631 &    Sy1.9    & FR I       & 44.15                         &   1.200        & 710      & 1,d   &  0.85 \\
SWIFT J1153.9+5848   & RGB J1153+585 &  0.2024 &    Sy1.5   & FR II       & 44.99                           &   3.298        & 255      & 2,e   &  0.84  \\
SWIFT J1238.4+5349   & RBS 1137      &  0.3475 &    Sy1     & FR II       & 45.50                           &   4.884        & 160      & 3,f  &  0.78  \\ 
SWIFT J1503.7+6850   & 4C+69.18       &  0.3180 &    Sy1.8   & FR II       & 45.41                           &   4.602        & 190      & 1,g  &  0.87 \\
\hline
IGR J13107--5626      & PMN J1310-5627 &  0.0930    &    Sy2     & FR II      & 44.56                         &   1.692        & 420    & 2,e   & 0.71 \\
\hline
SWIFT J0225.8+5946   & WB J0226+5927&  -    &    -        & FR II       & -10.99$^{\dagger}$              &    -           & 630      & 2e   &  -    \\
\hline
\hline
\end{tabular}
$^{\dagger}$ Since we are unable to estimate the source luminosity without a knowledge of its redshift, we report for completeness the logarithm value of the 14-195 keV flux  (in units of erg cm$^{-2}$ s$^{-1}$).\\
Previous GRG  reference: 1) \citet{2018ApJS..238....9K}  2) this work; 3) \citet{2020A&A...635A...5D}, 
4)  \citet{2016MNRAS.461.3165B}. \\
Radio image reference: a) \citet{2005AJ....130..896S}; b) \citet{1986A&A...169...63K}; c) \citet{2001MNRAS.326.1455M}; d) \citet{1999A&A...341...44S}; e) this work;  f) \citet{2011AJ....141...85R}; g) \citet{2001A&A...370..409L}
\end{threeparttable}
\label{tab1}
\end{table*}

\section{Results}
Following the release of new  INTEGRAL/Swift AGN surveys, we are now able to update the list of soft gamma-ray GRGs with respect to our previous work \citep{2016MNRAS.461.3165B}.
For INTEGRAL/IBIS, we consider  the  sample of  107 new AGN discussed by \citet{2016MNRAS.460...19M} with the addition of 
43  new objects reported by \citet{2016MNRAS.459..140M} in the deep extragalactic surveys
of M81, LMC and 3C273/Coma regions and those reported by \citet{2017MNRAS.470..512K} in the Galactic Plane Survey after 
14 years of INTEGRAL observations.
For Swift/BAT we use the 105 month catalogue of \citet{2018ApJS..235....4O} considering only the AGN of all types and the sources of Unknown class (U1, U2 and U3) listed among the  422 extra 
detections reported with respect to the 70 months catalogue. U type objects did not have a firm  optical counterpart at the time of the survey release due to the lack of X-ray data (U3) or having X-ray 
measurements but not an  X-ray detection (U2) or a credible optical (U1) association.

Following our previous work, we examined  the radio structure  of each 
new AGN  reported in the high energy catalogues or found through this work (see following discussion  and Appendix A),  by inspecting  images from the NRAO VLA Sky Survey (NVSS,  \citealt{1998AJ....115.1693C}), the VLA FIRST Survey (FIRST,  \citealt{1997ApJ...475..479W}), the  Sydney University Molonglo Sky Survey (SUMSS,
\citealt{2003MNRAS.342.1117M}),
or considering data already  available in the literature. 
For  objects located north of declination -40 degrees we used 1.4 GHz maps from the NVSS, the accuracy of which is $\sim$ 10$\arcsec$,
i.e. 1/4 of the survey angular resolution of 45$\arcsec$; the survey rms noise is 0.45 mJy/beam. We also complemented such information with images at the same frequency  from the FIRST survey whose
smaller point spread function or PSF (5$\arcsec$)
allows better accuracy (of the order of 1.5$\arcsec$) and an average rms noise  of 0.15 mJy/beam.
For particularly interesting sources (see following sections), we also inspected maps from the  recent VLA Sky Survey (VLASS Epoch 1, \citealt{2020PASP..132c5001L}) which provides a much better resolution (2.5$\arcsec$) at slightly higher frequencies and a typical rms noise of 0.12 mJy/beam.
For  objects located  south of
-40 degrees in declination,  we used 0.8 GHz SUMSS images; here the accuracy is worse, both in terms of angular resolution ($\sim$ 20$\arcsec$, as a result of a wider PSF) and rms noise (around 1 mJy/beam).

All together we inspected around  570 images to uncover those sources which display a double lobe morphology typical 
of radio galaxies of both FR I and FR II classes\footnote{FR I are sources whose luminosity decreases as the distance from the central galaxy or quasar host increases, while FR II sources exhibit increasing luminosity in the lobes 
\citep{1974MNRAS.167P..31F}}.

In this way, we extracted a list of 7 new soft gamma-ray selected GRGs, plus 1 object, SWIFT J0225.8+5946, which is large 
in extent (10.5 arcmin) but for which we still lack a redshift measurement, so that we cannot confirm at this stage 
if the source is a GRG or not. We note however that for any redshift above 0.06 the source will be classified as a GRG.

We further add to the list a previously reported radio galaxy, IGR J13107-5626 \citep{2016MNRAS.461.3165B}, for which we 
were able to obtain for the first time an optical spectrum and therefore estimate the source's class and redshift; 
details of the optical spectroscopy are reported in section 3.2 and in Appendix B. At the  measured redshift, the source can now be classified as a new giant radio galaxy, since  its size is just above 0.7 Mpc.

The new objects presented here and to be added to the sample  of GRGs  in  \citet{2016MNRAS.461.3165B} are all listed in Table~\ref{tab1}.
 All sources are new Swift/BAT gamma-ray detections with the exception of IGR J13107-5626 also seen by INTEGRAL/IBIS and   B2 1144+35B which instead is only listed in the INTEGRAL catalogue and  has  already been studied in some 
detail by our team   \citep{2019ApJ...875...88B,2020MNRAS.494..902B}.

Two objects in the sample, SWIFT J1153.9+5848 and SWIFT J1503.7+6850, were classified as Unknown (U1 and U3 respectively) in the BAT survey. 
We have been  able to update their class, and thanks to recent Swift/XRT observations,  found the X-ray/radio  counterpart 
of each of these two objects. In particular, we analysed all archival XRT pointings   and  stacked them together to enhance the signal  to noise ratio.  
The restricted soft X-ray error box  allowed the subsequent association of  both objects to radio galaxies of 
large extension: RGB J1153+585 and 4C+69.18 for SWIFT J1153.9+5848 and SWIFT J1503.7+6850 respectively 
(see Appendix A for details on the X-ray data analysis and results).

We also checked that for the remaining  7 objects, the X-ray/optical counterpart reported in the corresponding high energy  survey was the correct one.
For this purpose we examined available observations from the Swift/XRT archive for SWIFT J0632.1-5404, SWIFT J0636.5-2036,
SWIFT J0801.7+4764, B2 1144+35B, IGR J13107-5626 and SWIFT J0225.8+5946 and a Chandra pointing for SWIFT J1238.4+5349; 
the data analysis relative to all these objects is also reported in Appendix A.

All but one source in Table~\ref{tab1} display  an  FR  II  radio  morphology; the only exception is B2 1144+35B which belongs to the FR I class. 
The sample includes local AGNs (z$\le$0.35) of different optical classes with  type 1 and 2 objects being equally represented; 
we note that two objects have redshift above 0.3, i.e. slightly higher than those reported  by \citet{2016MNRAS.461.3165B} indicating that as soft 
gamma-ray surveys improve their  sensitivities (typically by a factor 1.4-1.5 with respect to our previous work),  GRGs at higher z are detected. 
The soft gamma-ray luminosities are within the range  observed by \citet{2016MNRAS.461.3165B} for the previous sample of 14 GRGs,
while  the radio sizes span from 0.7 to 1 Mpc.

\section{Newly discovered sources}
Most of the sources in the sample (see references in Table~\ref{tab1}) have already been discussed in the literature:  
in particular 5  are  listed in the catalogue of GRGs by \citet{2018ApJS..238....9K}, while one, SWIFT J1238.4+5349, 
has recently been reported as  a giant in the LOFAR Two-metre Sky Survey (first data release) by \citet{2020A&A...635A...5D}.
IGR J13107-5626 was suggested as a candidate giant radio galaxy by \citet{ 2016MNRAS.461.3165B}, and we can now confirm 
this initial suggestion. For the remaining  object, SWIFT J1153.9+5848, no mention  to its large radio structure was found 
in the literature. Therefore we conclude that  two sources in the sample are newly discovered GRGs (SWIFT J1153.9+5848 and
IGRJ13107-5626) while one is presented here as a candidate giant radio source (SWIFT J0225.8+5946), waiting for a redshift
estimate to confirm its optical class and radio typology. In the following section, each of these three sources will be discussed
individually in an attempt to provide further information on their nature.

\subsection {SWIFT J1153.9+5848/RGB J1153+585}
SWIFT J1153.9+5848 has been associated and classified through this work for the first time;  the optical counterpart 
RGB J1153+585  has mixed classification in NED and SIMBAD, being listed either as a Seyfert 1.5 or a BL Lac.
We adopt here the Seyfert classification since the SDSS spectrum  shows emission lines and in particular a broad H$\alpha$
line,  all at a redshift of 0.202 (see \citealt{2012MNRAS.423..600S} for details of the optical spectrum).

\begin{figure*}
\begin{center}
\includegraphics[width=\textwidth]{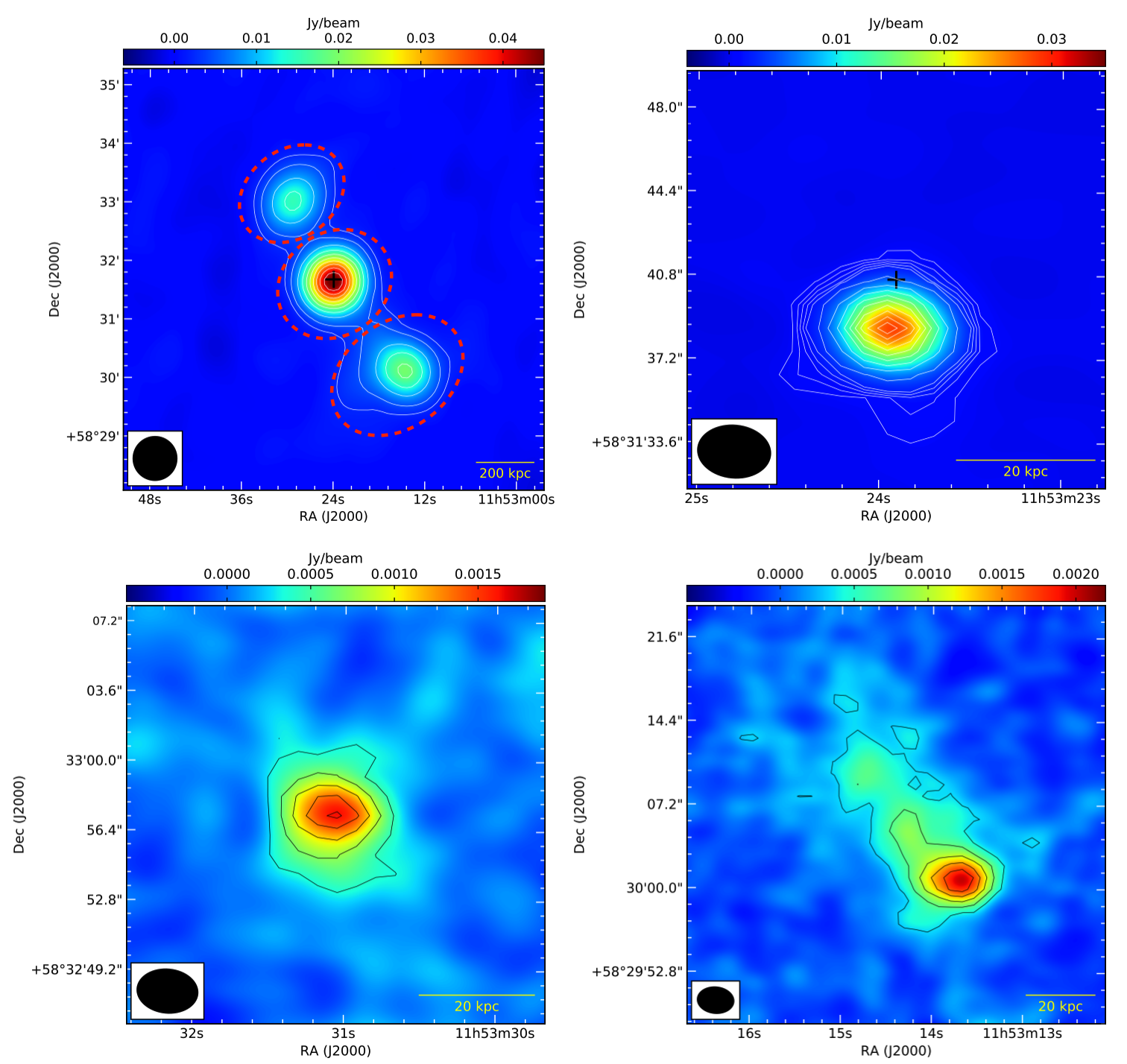}
\end{center}
\caption{Images from NVSS (entire source in the top left panel) and VLASS (core in the top right panel plus  north-eastern and  south-western lobes in the bottom left and right panel respectively) for SWIFT J1153.9+5848/RGB J1153+585.
The NVSS  contours are 5,10,20,30,40,50,60,70,80,90$\%$ of the source peak flux density which is 46.7mJy/beam. The red dashed regions show the areas used for the flux density extraction. 
The VLASS contours are 1,2,3,4,5,10,20,30,40,50,60,70,80,90$\%$ of the peak flux density which is 35.8 mJy/beam.
The black cross in the top figures marks the position of the X-ray core.}
\label{fig1}
\end{figure*}

The radio image of the entire source as reported in the NVSS  is shown in figure \ref{fig1} (top left panel): the radio morphology is typical of an FR II galaxy with a  bright  core marked by the X-ray position and two prominent and fairly symmetric lobes. 
The NVSS core and total lobe flux densities are 51.8 and 47.6 mJy respectively; the total lobe flux is  almost equally divided between the north-eastern  (23.3 mJy) and south-western (24.3 mJy) component.

Also shown in figure \ref{fig1} are zoomed images of  each source component (core in the top right panel and individual lobes in the bottom left and right panels) as seen by the VLASS.
The core flux density at 3 GHz\footnote{VLASS observational bandwidth is 2-4 GHz, so that the middle of this, i.e. 3 GHz,  is taken as the reference frequency}  is 37.5 $\pm$ 3.7 mJy, while that from  the north-eastern  and south-western  lobe is of   4.4 and 13.9 mJy respectively (10$\%$ uncertainty).
The slight misplacement (2.2 $\arcsec$) between X-ray and radio core  is perfectly compatible with the X-ray  positional uncertainty; the more accurate VLASS position locates the core of this source at  RA(J2000)= 11 53 23.92 and Dec(J2000)= +58 31 38.479 (0.5 $\arcsec$ uncertainty). 

Comparison between  NVSS and FIRST surveys indicates that the core flux density is variable over long (yearly) timescale: 51.8 versus 79.4 mJy, respectively \citep{2018A&C....25..176A}. The lobe flux densities  measured by FIRST (5.8 and 14.6 mJy for the north-eastern  and south-western lobe respectively with a 10$\%$ uncertainty) are similar to those of VLASS, implying a flatter spectrum than typically seen in the lobes as expected when probing their most compact and brightest part (i.e. the hot spots visible in figure \ref{fig1}) with high-resolution interferometric observations. 

The flux densities provided by NVSS/FIRST and VLASS surveys, allow also to  estimate a range of values for the spectral index\footnote{Here and in the following, we define the radio spectral index through the expression for flux density 
as a function of frequency: S($\nu)=\nu^{\alpha}$. } of the core (-0.42/-0.99), which is clearly  influenced by the source variability.

Finally from the flux density  values we can also estimate the  source radio core dominance,  defined here and throughout the paper as R =S$_{Core}$/(S$_{Tot}$ - S$_{Core}$), where S$_{Core}$ and S$_{Tot}$ are the core and the total flux densities at a given frequency, respectively.
For most of our sources, R can be approximated by the relative core to lobe flux density ratio  with the only exception of PKS 0634-20, where the total source emission (estimated to be 8.48 Jy from the NVSS) is twice as large as that of the lobes alone (see section 4.2). 
The  core dominance of  SWIFT J1153.9+5848  estimated from the NVSS data is 1. 

SWIFT J1153.9+5848 is also listed in the VLBA Imaging and Polarimetry Survey
(VIPS, \citealt{2007ApJ...658..203H})  where it is reported  as a core jet source (jet shorter than 6 mas) 
with a 5 GHz flux density of 51.8 mJy. 
Furthermore, the source is among a set of soft gamma-ray selected GRGs which have LOFAR data and  
will therefore be the target of a dedicated study (Bruni et al. in preparation).\\
Beside being  core dominated in radio, this GRG  is also  quite  bright at  X/soft gamma-ray energies with a 
2-10 kev and 14-195 keV  luminosities around 9.4 $\times$ 10$^{43}$ erg s$^{-1}$ and 10$^{45}$ erg s$^{-1}$
respectively. 
If we evaluate the bolometric luminosity, using the correction of \citet{2004MNRAS.351..169M} for the 2-10 keV 
luminosity (see section 4) we obtain a value of 3.1 $\times$ 10$^{45}$  erg s$^{-1}$. 
Assuming instead the relation L$_{Bol}$ = 15 $\times$ L$_{\textrm{14-195 keV}}$, found by 
\citet{2008ApJ...684L..65M}, we obtain a value of 1.5 $\times$ 10$^{46}$ erg s$^{-1}$. 
For a black hole mass of  2.5$\times$ 10$^{8}$  solar masses \citep{2007ApJ...667..131G}, 
the Eddington luminosity is  3.2 $\times$ 10$^{46}$ erg s$^{-1}$, implying an Eddington ratio around 0.1 - 0.5, resulting in the most
efficient accretor in the current sample of GRGs. \\
This radio galaxy is also mildly variable at  optical wavelength \citep{2019AN....340..437A}; evidence of 
multi-waveband variability, high core dominance and high accretion rate suggest a blazar like core. 
Indeed the source is reported as a flat spectrum radio QSO by \citet{2005ApJ...626...95S} and is also listed 
in the Rome blazar (BZ) catalogue by \citet{2009A&A...495..691M} with the same classification. 
This blazar like core, which seems very active  at the moment, is however embedded in a large double lobe
structure which is presumably very old as indicated  by the  time it took to reach its current size (around 25 Myr
assuming a typical velocity of advancement of 0.1c);  thus the  source
may have restarted activity only recently.
Further radio observations  at low and high angular scales will shed light on this issue.

\begin{figure}
\includegraphics[width=\columnwidth]{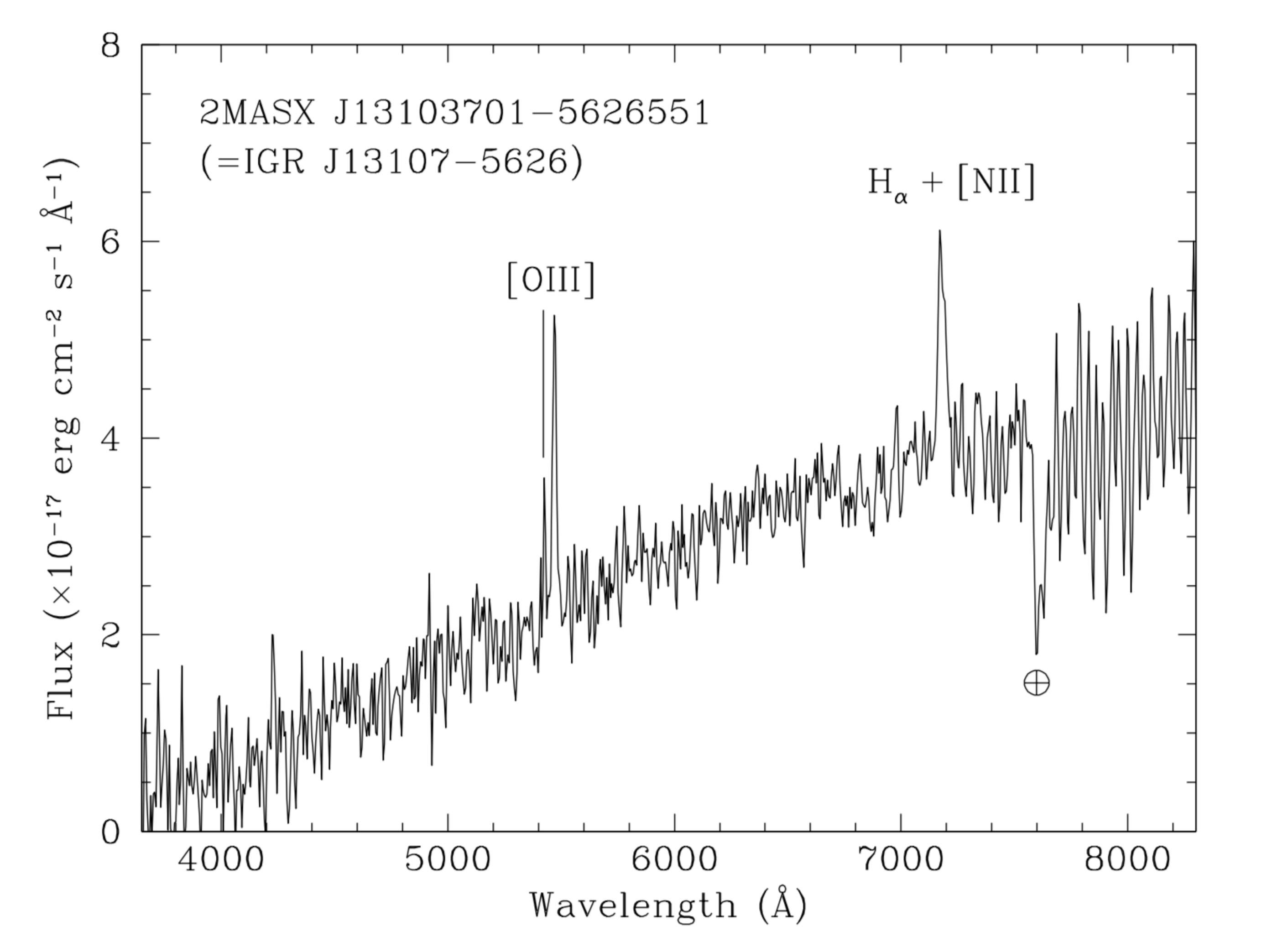}
 \caption{Optical spectrum (not corrected for the intervening Galactic absorption) of the counterpart of IGR J13107-5626.
The main spectral features are labelled. The symbol $\oplus$ 
indicates a telluric absorption band due to atmospheric O$_2$.}
    \label{fig2}
\end{figure}

\subsection {IGR J13107--5626/PMN J1310--5627}
This source was the only one in \citet{2016MNRAS.461.3165B} sample with no redshift, so that it was impossible to estimate 
the source size at the time of the paper. However the large linear extent of the source suggested a large size, which prompted a
follow up study in order to estimate the source distance and optical class; in particular we analyse optical spectra of the source available at the ESO archive (see Appendix B for details of the optical data analysis).

The analysis of the  source spectrum (see figure \ref{fig2}) shows the presence of two groups of narrow emission
lines superimposed on a reddened (and likely absorbed) continuum; 
we identify these features as [O {\sc iii}] $\lambda$5007 and as
the H$_\alpha$+[N {\sc ii}] complex at an average common redshift 
$z$ = 0.093 $\pm$ 0.001. The physical properties of the main emission lines in the optical spectrum are reported in Table~\ref{tab2}.

\begin{table}
\caption{Physical properties of the main emission lines in the optical
spectrum of the counterpart of IGR J13107-5626. Errors are at 1$\sigma$ 
confidence level, whereas upper limits are at 3$\sigma$.}
\begin{tabular}{lrrc}
\hline
\hline
Line                 &      Flux$^*$   &  $\lambda_{\rm obs}$  &   $z$ \\     
\hline
[O {\sc iii}]        & 6.7$\pm$0.7 & 5470$\pm$3 & 0.092$\pm$0.001 \\
H$_{\alpha}$         & 4.9$\pm$1.0 & 7172$\pm$3 & 0.093$\pm$0.001 \\
\Nii                   & 3.3$\pm$0.7 & 7195$\pm$3 & 0.093$\pm$0.001 \\
\hline
H$_\beta$        & $<$1        &  --- & --- \\
\hline
\multicolumn{4}{l}{$^*$in units of 10$^{-16}$ erg cm$^{-2}$ s$^{-1}$} \\
\hline
\hline
\end{tabular}
\label{tab2}
\end{table}

The relative strengths of [N {\sc ii}]
and H$_{\alpha}$ (their observed ratio is 0.67$\pm$0.09), together 
with the absence of H$_{\beta}$, point to a Type 2 AGN classification for this object, thus securing its identification 
as the optical counterpart of the high-energy source IGR J13107-5626.
The source X-ray spectrum is as expected absorbed (see also \citealt{2010MNRAS.403..945L}) and the source 2-10 keV  
bolometric luminosity is 3.4 $\times$ 10$^{45}$ erg s$^{-1}$ using the correction of \citet{2004MNRAS.351..169M}; 
assuming instead the relation of \citet{2008ApJ...684L..65M}, we obtain a value of 5.4 $\times$ 10$^{45}$ erg s$^{-1}$ 
for the 14-195 keV bolometric luminosity.
Unfortunately the black hole mass is not yet available to estimate the source Eddington ratio with some accuracy; however assuming a mass in the range  10$^{8}$-10$^{10}$ solar masses (as seen in all our GRGs) we obtain Eddington ratios in the range 0.003-0.4. 

\begin{figure}
\includegraphics[width=\columnwidth]{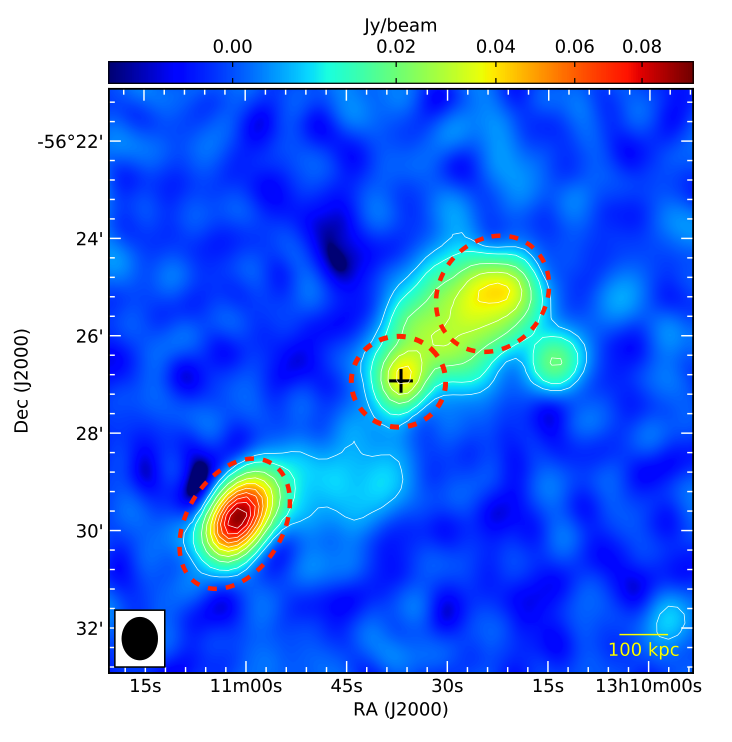}
 \caption{SUMSS image of IGR J13107--5626/PMN J1310--5627.
It shows the core and two lobes. 
The SUMSS countours are 5,10,20,30,40,50,60,70,80,90 $\%$ of the peak flux density which is 97.3 mJy/beam. The red dashed regions show the areas used for the flux density extraction. The black cross  marks the position of the X-ray core.}
\label{fig3}
\end{figure}

At radio frequencies the source is poorly studied, but  southern surveys provide some useful information.  
The radio image of  PMN J1310--5627 as reported in 
the SUMSS is shown in figure \ref{fig3}: the radio morphology is typical of an FR II galaxy with a core and two prominent lobes. The radio core which coincides with the X-ray position, is attached to the north-western lobe but it  is detached from the south-eastern one; there is some off axis emission  south of the core which may be explained as a hydro-dynamical back-flow, i.e. plasma deflected by the host galaxy halo.

The core flux density at 0.8 GHz is 55.6 mJy, while that from  the  north-western and south-eastern lobe 
is  82 and 169 mJy respectively (10$\%$ uncertainty). This indicates a core dominance of  0.2. 
The source core  is also listed in the catalogue of PMN objects observed with the Australia Telescope Compact Array
\citep{2012MNRAS.422.1527M}, where it is reported with a  4.8 and 8.6 GHz  flux density of 70 and 53  mJy respectively (spectral index  -0.5). 

This source is  poorly studied at all wavelengths but deserves 
to be the target of a dedicated observational campaign in order to  compare its
properties, especially in terms of restarting activity, with other GRGs in our sample.

\subsection {SWIFT J0225.8+5946/WB J0226+5927}
Finally SWIFT J0225.8+5946 is a GRG if its  redshift exceeds  0.06; unfortunately the source is quite dim at optical wavelengths (B,R magnitudes  fainter than 20); indeed it is 
listed  in the  Pan-STARRS1 (PS1) catalogue with i and z magnitudes of 21.63 and 20.66 respectively  \citep{2016arXiv161205560C}. Therefore a telescope with diameter of at least 3-4 meters is needed to obtain a redshift measurement; for the moment we therefore consider this object  as a candidate GRG. 

The source WISE colours (W1-W2=1.18, W2-W3=2.85) indicate that the core of this source is an AGN \citep{2015ApJS..221...12S};
furthermore these colours suggest  that the source may be similar to a blazar of intermediate type between a BL Lac and a FSRQ 
\citep{2014ApJS..215...14D}. Interestingly  the source is absorbed at X-ray frequencies which may lead to a possible  type 2 optical classification.

\begin{figure*}
\begin{center}
\includegraphics[width=\textwidth]{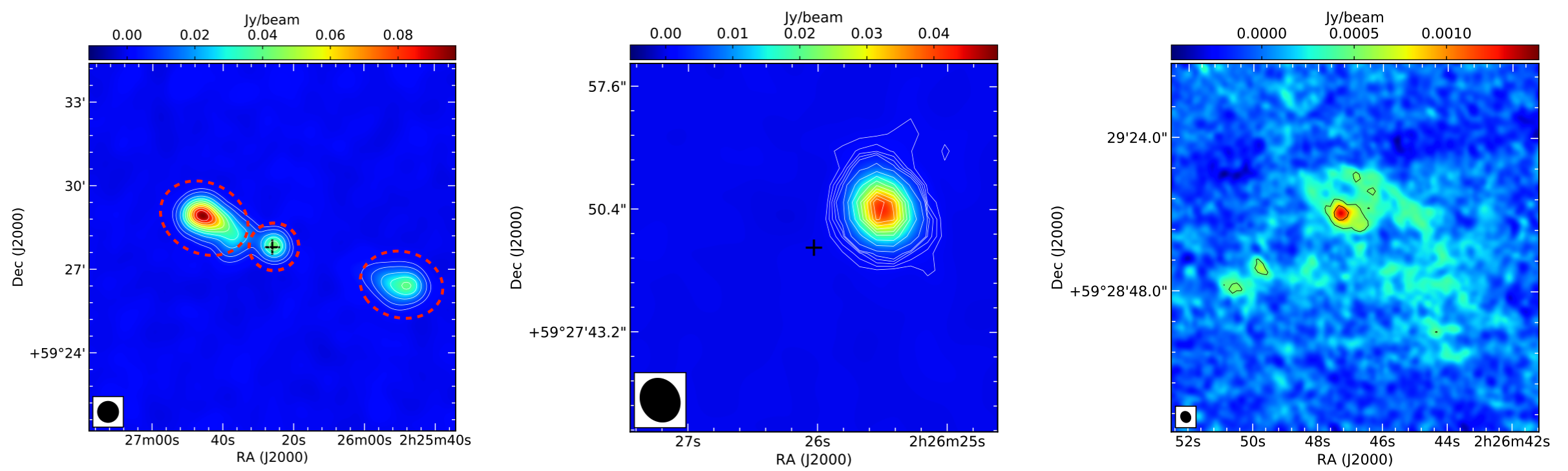}
\end{center}
\caption{Images from NVSS (entire source in the left panel) and VLASS (core in the middle  panel plus  north-eastern lobe  in the right panel) of SWIFT J0225.8+5946/WB J0226+5927. The NVSS  contours are 5,10,20,30,40,50,60,70,80,90$\%$ of the source peak flux density which is 96.7 mJy/beam. The red dashed regions show the areas used for the flux density extraction.
The VLASS contours are  1,2,3,4,5,10,20,30,40,50,60,70,80,90$\%$ of the peak flux density which is 49.5 mJy/beam.
The black cross in the top left and middle panels  marks the position of the X-ray core.}
\label{fig4}
\end{figure*}

The source radio image from the NVSS  is shown in figure \ref{fig4} (left panel), 
where it is evident the large extent of the source and a clear FRII morphology; the core is attached to the north-eastern lobe whereas the south-western one  seems to be detached from the rest of the source.
The core flux density at 1.4 GHz is 50 mJy, while those from  the north-eastern and south-western  lobes are of  234 and 94 mJy respectively
(10$\%$ uncertainty). This indicates a core dominance of  0.15.
The source is also listed in the Canadian Galactic Plane Survey \citep{2017AJ....153..113T} with a similar 1.4 GHz core flux density.

Also shown in figure \ref{fig4} are the VLASS images of the core (middle panel) and north-eastern lobe (right panel)  while the south-western one is not detected. The core flux density at 3 GHz  is 58$\pm$9 mJy while that of the only visible lobe is 48.8 mJy. The more precise positional  accuracy of VLASS  locates the source core  at RA(J2000)=  02 26 25.53 and Dec(J2000)= +59 27 50.933 (uncertainty around 0.5$\arcsec$), only 2.8$\arcsec$ away from the X-ray position.

The source core  is also reported in the AMI Galactic plane survey \citep{2013MNRAS.429.3330P} with a 16 GHz flux density of
61.7$\pm$6.6 mJy, suggesting a flat spectrum. This again is a poorly studied source, but before performing a more accurate  analysis, it is mandatory to obtain a redshift estimate to confirm its GRG nature. 

\begin{figure}
\includegraphics[width=\columnwidth]{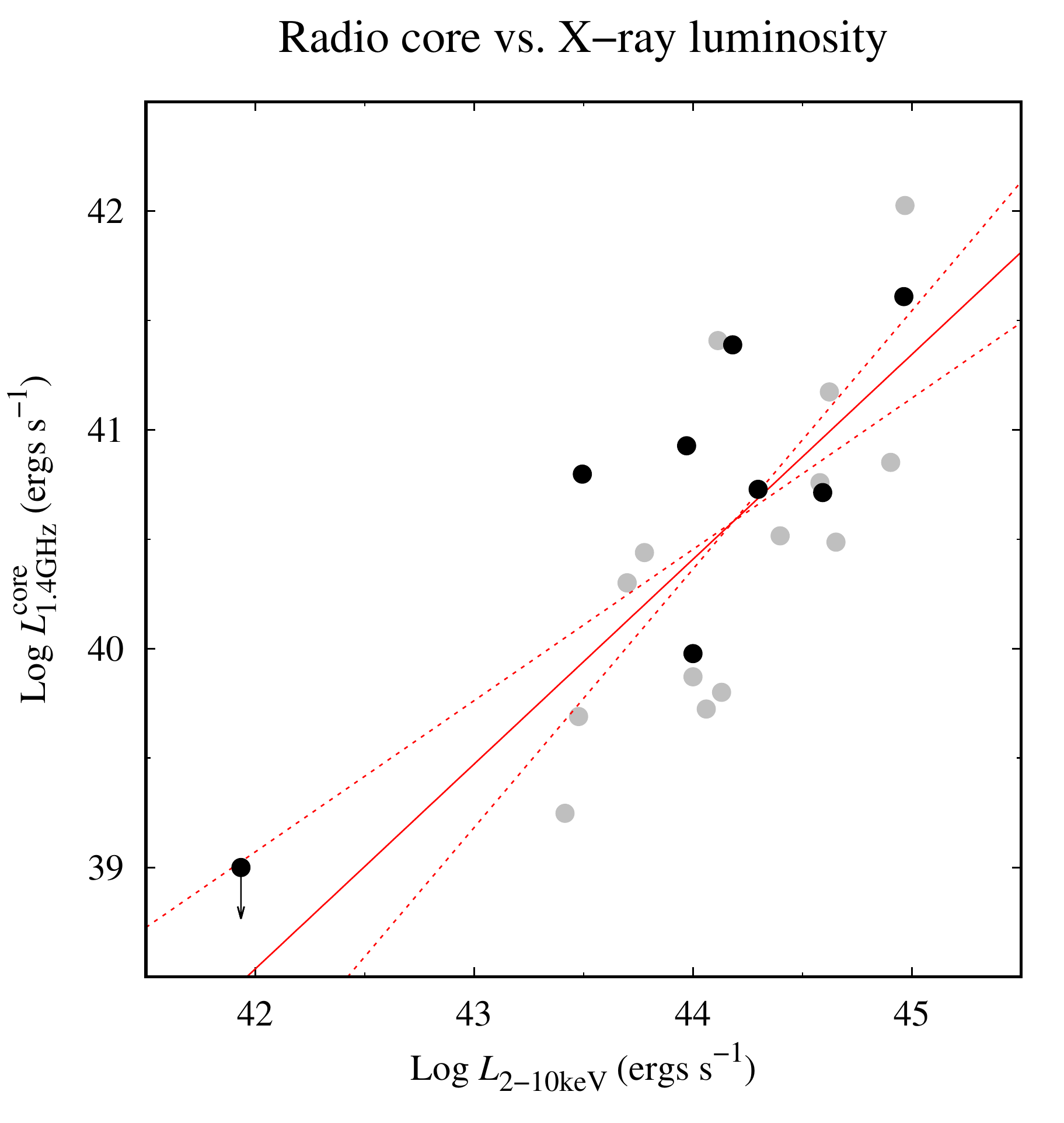}
 \caption{Radio core 1.4-GHz luminosity versus X-ray 2--10 keV luminosity. The red solid line represents a linear fit in the log-log space, while the red dashed lines correspond to the 90 per cent error on the slope and normalization.The new GRGs are in black while the old ones are in gray.The only source with an upper limit on the core 1.4-GHz luminosity is PKS 0634-20.}
\label{fig5}
\end{figure}

\begin{figure}
\includegraphics[width=\columnwidth]{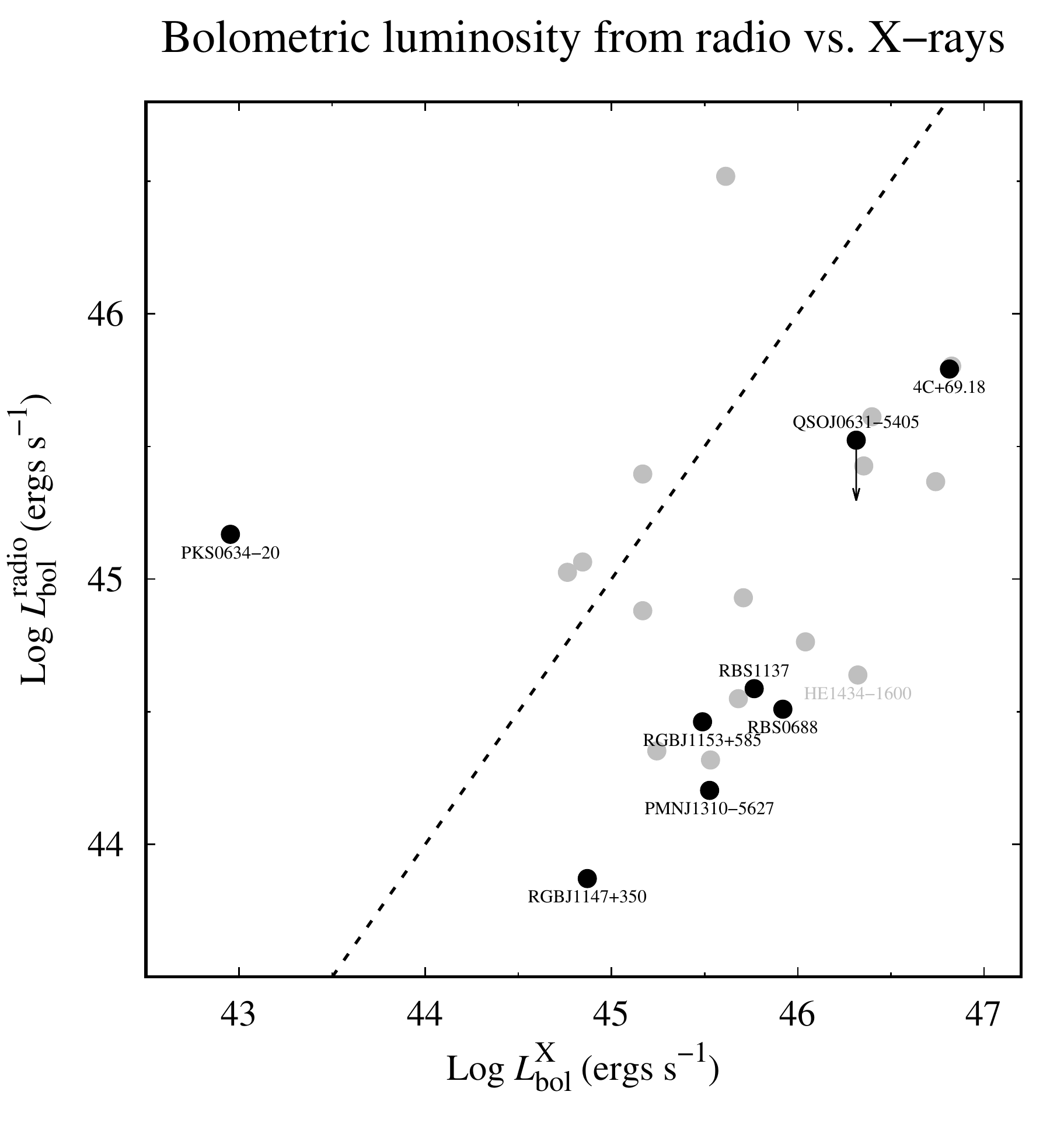}
 \caption{Bolometric luminosity estimated from the radio luminosity of the lobes (y axis) versus that estimated from the 2--10 keV luminosity (x axis).
The dashed line represents the identity y = x.The new GRGs are in black while the old ones are in gray. In this figure we also correct a mistake done in \citet{2018MNRAS.481.4250U}  that caused a wrong placement of HE 1434-1600 (see their Fig. 2).} 
    \label{fig6} 
\end{figure}

\begin{table*}
\footnotesize
\begin{threeparttable}[b]
\caption{X-ray and radio  properties of the new soft $\gamma$-ray selected GRGs.
\label{tab3}
From column 1 to 10 we list the source name from SIMBAD, the core and lobe radio  flux densities in mJy, the core and lobe radio luminosities in erg s$^{-1}$, the core dominance R with  relative reference, the 2--10 keV X-ray luminosity in erg s$^{-1}$, the log of the  black hole mass in solar masses and the Eddington ratios.}
\centering
\begin{tabular}{l c c c c c c c c c}
\hline\hline
Name              & $F_{Core}$$^{\dagger}$  & $F_{Lobe}$$^{\dagger}$    & $L_{Core}$$^{\dagger}$              & $L_{Lobe}$$^{\dagger}$        & R  (ref) &  $L_{2-10keV}$  & Log$M_{BH}$,(ref) & $\lambda_{1}$/$\lambda_{2}$$^{\ddagger}$\\

                  &   mJy                 &   mJy                 & 10$^{40}$ erg s$^{-1}$         & 10$^{41}$ erg s$^{-1}$   &           &
                  10$^{43}$ erg s$^{-1}$ &  &  \\

\hline\hline
QSO J0631-5405$^{a}$    &    37.7    &   $<$ 656     &  5.17             & $<$8.99           &   $>$ 0.060,  (1)      &   39.14      &   9.5, (a) & 0.06/0.03  \\  
PKS 0634-20       &  $<$10.0   &   3970            &  $<$0.1          & 3.97           &   $<$0.001, (2,3)      &   0.086      &   8.2, (b) & $3 \times 10^{-4}$/0.05 \\
RBS 0688            &   57.7     &    94           &  5.35              &  0.87             &    0.613, (4)          &   19.85      &   8.9, (c)  & 0.09/0.08\\
RGB J1147+350     &   340-609  &    148.4     &  4.50-8.06         &  0.20             &   2.3-4.1,   (5)      &    3.12      &   8.2, (d) &  0.04/0.11 \\
RGB J1153+585     &   51.8     &    47.6         &  8.46              &  0.78              &    1.088,   (3)       &    9.35      &   8.4, (e) & 0.1/0.47 \\
RBS 1137         &   43.2    &     18.4          & 24.47              &  1.04              &    2.340, (6,7)       &   15.19      &   9.7, (c) & 0.01/0.07  \\
4C +69.18         &   88.4     &    362.6           & 40.64              &  16.67             &    0.244,  (8)        &   91.95      &   --    &    --    \\
\hline
PMN J1310-5627$^{b}$    &   55.6      &    251            &  0.95              &  0.43               &    0.220(3)                &    10      &   --   &--       \\
\hline   
WB J0226+5927     &   50     &    328          &  --                &  --                &    0.152,  (3)        &     --       &   -- &--         \\
\hline

\end{tabular}
$^{\dagger}$ All radio flux densities  are at 1.4GHz, unless otherwise stated\\ 
$^{a}$ For this source the lobe flux density is given at 0.8GHz; however assuming that lobe spectral indeces are generally steep, we can assume this flux density  to be also an upper  limit to the 1.4 GHz lobe flux density\\
$^{b}$ For this source both core and lobe flux densities  are given at 0.8 GHz, consequently the core dominance refer to this frequency.\\
$^{\ddagger}$ $\lambda_{1}$ is estimated using the  bolometric correction of \citet{2004MNRAS.351..169M} for the 2--10 keV luminosity while  $\lambda_{2}$ is estimated using the  bolometric correction of \citet{2008ApJ...684L..65M} 
for the 14--195 keV luminosity\\
Core dominance references: 1) \citet{2005AJ....130..896S}; 2) \citet{2000ApJS..131...95F}; 3) this work using NVSS survey for PKS 0634-20,  RGB 1153+585 and WB J0226+5927 and SUMSS survey for PMN J1310-5627; 4) \citet{2012MNRAS.426..851K}; 5) \citet{1999A&A...341...44S}; 6) \citet{2009AJ....137...42R}; 7) \citet{2011AJ....141...85R}; 8) \citet{2001A&A...370..409L}\\
Black hole mass references: a) \citet{2003MNRAS.339.1081D}; b) \citet{2002ApJ...579..530W}; 
c) \citet{2015MNRAS.454.3864B}; d) \citet{2009MNRAS.398.1905W}; e) \citet{2007ApJ...667..131G}\\
\end{threeparttable}
\end{table*}

\section{Discussion}

\subsection{Radio and X-ray properties}
To explore further the radio properties of our sources and to compare them with the old GRG sample, we collected 
from the literature or estimated in this work the  radio flux densities and luminosities, mostly relying 
on high-resolution  0.8-1.4  GHz (SUMSS-NVSS)  data  that  allow to   disentangle  the 
different contributions from the core and the lobes.   
In this way we also were able to estimate the core dominance for all objects.
Finally we report when possible the central  black hole mass of the AGN. 
These parameters are all listed  in Table~\ref{tab3}.
The core dominance spans a large range of values from 0.03 
to greater than 1,  while the black hole masses are also quite high (greater than 10$^{8}$ solar masses).

Following \citet{2018MNRAS.481.4250U} we also compared the X-ray and radio luminosities of the core region, adding  8 new sources to our previous sample of GRGs (see Table~\ref{tab3} and figure \ref{fig5}). As examined in detail  by Ursini and co-workers, the major source of uncertainty on the parameters reported in  figure \ref{fig5} is flux variability (which can be roughly of a factor of 2-3), while the measurement errors are relatively small (only a few percent).
The Kendall's correlation coefficient for the whole set of sources is 0.44, with a p-value of 5e-3.
For simplicity, to perform the linear fit and compute the correlation we did not take into account PKS 0634-20, for which we only have an upper limit to the radio core flux density (which is anyway fully compatible with the linear correlation as evident in the figure).
With respect to our previous work, the  correlation between the two
parameters becomes slightly flatter with coefficient 0.9$\pm$0.2  instead of 1.1$\pm$0.3, but  still within uncertainties
(see figure \ref{fig5}). As extensively discussed by \citet{2018MNRAS.481.4250U} there are two different branches in the radio--X-ray
correlation originally found in X-ray binaries and later extended to AGN: a  standard branch with coefficient close to 0.6 and a  second
one with coefficient in the range 1--1.4. The first branch is consistent with the source being powered by a radiatively  inefficient accretion flow while the  second one is  thought to be associated  with a radiatively efficient flow. Therefore a coefficient close to 1
would locate soft gamma-ray  selected GRGs on the 'efficient' branch of the radio--X-ray correlation diagram, implying a radiatively efficient mode of accretion for these sources. 

Following again \citet{2018MNRAS.481.4250U}, we also calculated two independent estimates of the bolometric luminosity, 
one from the core X-ray luminosity (which we label $L_{\textrm{bol}}^{\textrm{X}}$)
and the other from  the lobe 
radio luminosity (which we label $L_{\textrm{bol}}^{\textrm{radio}}$)
 
The first estimate,  $L_{\textrm{bol}}^{\textrm{X}}$), is based on the bolometric correction of \citet{2004MNRAS.351..169M}
for the 2--10 keV luminosity (see \citealt{2018MNRAS.481.4250U} for details).  As a byproduct of this estimate, we can also
evaluate the Eddington ratio  $\lambda_{1}$,  for all those sources having a measurement of the black hole mass; similarly we estimate the Eddington ratio 
$\lambda_{2}$ using the bolometric  correction of \citet{2008ApJ...684L..65M} for the 14-195 keV luminosity (see last column of Table \ref{tab3} for both $\lambda_{1}$ and $\lambda_{2}$ values). It is evident from the table that most objects are, as indicated by the correlation shown in figure \ref{fig5},  efficient accretors with Eddington ratios above 0.01, with the possible exception of PKS 0634-20 which has a much  lower accretion rate ($3 \times 10^{-4}$) based on the 2-10 keV luminosity (but see the considerations made in the next section).  

The second estimate of the bolometric luminosity is based on a correlation with the 1.4-GHz lobes luminosity reported by
\citet{2015MNRAS.446.2985V}, which can be written in the form:

\begin{equation}
\log L_{\textrm{bol}}^{\textrm{radio}} = \log L_{\textrm{1.4GHz}}^{\textrm{lobes}} + 3.57.
\end{equation}

Comparison between  the above  two estimates provides a simple way to relate the AGN activity as traced by the X-ray emission of the nucleus with the one traced by the luminosity of the radio lobes. 
Because the scatter of the \citet{2004MNRAS.351..169M}
relation is $\sim$0.1 dex, the main source of uncertainty on $L_{\textrm{bol}}^{\textrm{X}}$  is the
X-ray flux variability while that on $L_{\textrm{bol}}^{\textrm{radio}}$ is the  scatter of the 
Van Velzen et al. relation  which amounts to 0.3--0.5 dex.
Also in this case and despite the uncertainties, we confirm the general trend observed previously
that the bolometric luminosity given by the radio lobes is relatively low (around one order of magnitude) compared to the ones given by the nuclear X-ray emission, with
only one exception (see figure~\ref{fig6}).
As discussed by \citet{2018MNRAS.481.4250U}, this indicates that either the nucleus is now more powerful than in the past,
consistent with a restarting of the central engine, or that the giant lobes are dimmer due to expansion losses. The finding that a large fraction of soft gamma-rayselected GRGs host a young radio nucleus or have indication of multiple
activity episodes from their radio morphology  \citep{2019ApJ...875...88B, 2020MNRAS.494..902B} suggests that also for these new objects  a scenario of restarting activity is the most likely one.

\subsection{The peculiar case of SWIFT J0636.5-2036/PKS 0634-20}

The only exception to the above  picture is SWIFT J0636.5-2036/PKS 0634-20, which in figure~\ref{fig6} is located in a region of its own. This location could reflect a wrong estimate  either of the X-ray/core bolometric luminosity or of the radio/lobe bolometric one. 
The first could be due to  heavy X-ray obscuration present in the source but  not properly accounted for due to the poor quality of the 
Swift/XRT  spectrum: the source is in fact a type 2 AGN and some intrinsic absorption is expected, while the XRT signal to noise ratio is the lowest among
the sources analysed here (see Table in  Appendix A)are the core and the total flux densitie
Indeed \citet{1995ApJ...454..683S} reported strong visual extinction in the source of roughly 30 magnitudes, which implies,
for a normal gas  to dust ratio\footnote{$N_{H}$=$A_{V}$/2 $\times$ 10$^{21}$ at cm$^{-2}$}, an X-ray column density of at least a few  10$^{22}$ at cm$^{-2}$; this  should be considered as a lower limit as generally extinction in X-rays  
is stronger than at optical/IR wavelengths. Indeed an old ASCA spectrum of the source, albeit of low  signal to noise ratio,
indicates that the column density could be as high as 8 $\times$ 10$^{23}$ at cm$^{-2}$  \citep{1999ApJ...526...60S} with a
consequent large correction (around a factor of 35) to the observed 2-10 keV luminosity reported in Table~\ref{tab3}.
This correction is close to the value needed to bring the source more in line with all other GRGs  indicating  that only a
heavily absorbed, possibly  Compton thick source could explain the peculiar location in figure~\ref{fig6}. The Compton thick nature of
PKS~0634-20 could also be assessed with several diagnostic indicators: for example \citet{1999ApJS..121..473B} proposed the
X-ray/OIII flux ratio while more recently  \citet{2007ApJ...668...81M} proposed the  $F_{2-10keV}$/ $F_{2-100keV}$ softness
ratio, where both X-ray fluxes are the observed ones, i.e. not corrected for absorption.
In the case of PKS~0634-20 these two diagnostics provide  values of 0.33  and 0.02  respectively; both values locate the
source in the region populated by Compton thick objects indicating an X-ray column density around 10$^{24}$ at cm$^{-2}$,
i.e. not far from the value measured by ASCA. Incidentally heavy intrinsic absorption in X-rays could also alleviate the problem of a
low $\lambda_{1}$ value while filling at the same time the gap with the much higher $\lambda_{2}$ estimate.
Only a dedicated long X-ray observation for example combining  XMM/NuSTAR pointings can confirm this indication. 

\begin{figure*}
\begin{center}
\includegraphics[width=\textwidth]{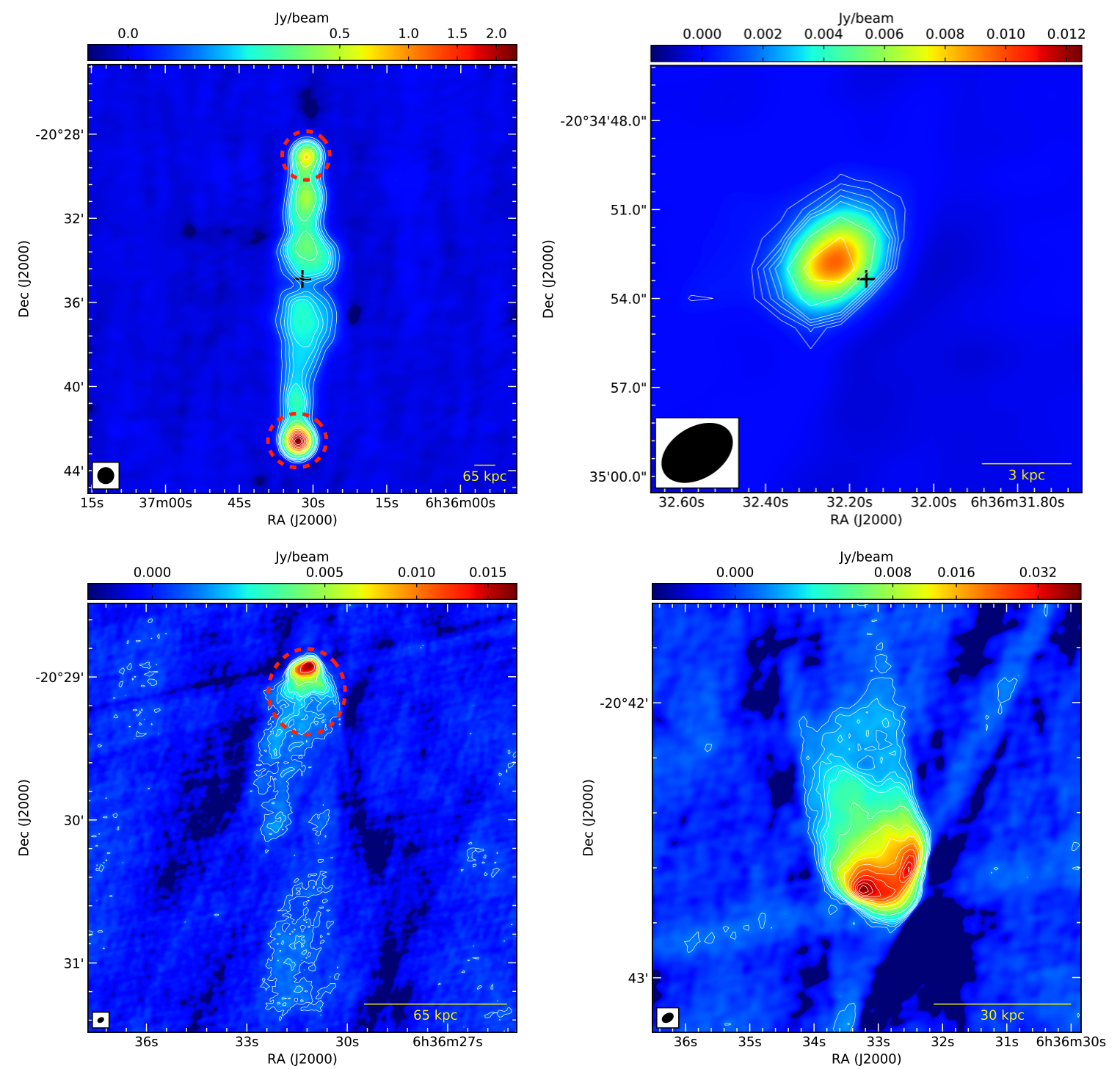}
\end{center}
\caption{Images from NVSS (entire source in the top left panel) and VLASS (core in the top right panel plus  north and  south lobes in the bottom left and right panel respectively) for SWIFT J0636.5-2036/PKS 0634-20.
Contours are 1,2,3,4,5,10,20,30,40,50,60,70,80,90$\%$ of the source peak flux density which is 2300 and 43.4 mJy/beam for the NVSS and VLASS respectively. The red dashed regions show the areas used for the flux density extraction.  The black cross in the top figures marks the position of the X-ray core.}
\label{fig7}
\end{figure*}

Alternatively the radio flux density from the source lobes could be overestimated. The source displays a  complex morphology, with a core,  two lobes and emission in between (see NVSS image in figure~\ref{fig7}, top left panel).
The core X-ray position falls in a region of minimum emission so that  only  an upper limit to the 1.4GHz flux density is available; furthermore between lobes and core a broadened emission region is visible,
resembling a plasma back-flow. Separating these  various components  at the angular resolution available at 1.4GHz  is not  straightforward and it  is   possible  that  the lobe flux density estimated in this work and reported in Table \ref{tab3} is not  perfectly quantified. Indeed the regions used to  extract the lobes flux density  have been defined in a conservative way, stopping  in correspondence of the first minimum of emission
found on the lobe-core axis (see red circles in the top left panel of figure ~\ref{fig7}).

To provide an alternative view  at  nearby frequencies, we inspected the VLASS image of the source 
(see also figure~\ref{fig7}) and evaluated the flux density from each component.
The more accurate VLASS  coordinates locate the source core at  RA(J2000)= 06 36 32.21   and Dec(J2000)= -20 34 52.843 (uncertainty 0.5 $\arcsec$), i.e.very close to the X-ray position.
The core flux density at 3 GHz is 15 mJy, while that from  
the northern and southern lobe is of  342 and 1290 mJy respectively (10$\%$ uncertainty). 
The difference between these values and those reported in  Table~\ref{tab3} for the lobes (quantified in a factor of $\sim$ 2) could be due to many factors. First of all,  NVSS and VLASS   observations were taken at different frequencies,   angular resolution and VLA configurations; moreover it is likely  that the oldest electron populations contribute less to the lobe  radio emission at higher frequencies, thus  providing less flux density  in the VLASS. The overall result is that only the most compact emission from the lobe is observed at 3 GHz. To conclude, the NVSS total lobe flux density is, at most, only  slightly overestimated.

Taking into account all the above considerations, we conclude that  the source position in figure~\ref{fig6} can be moved  closer to  the line  of equality between core and lobes bolometric luminosities (especially if the source is heavily absorbed in X-rays), solving its  peculiarity with respect to the other GRGs.

\section{Conclusions}
A set of new soft gamma-ray selected GRGs has been
extracted from recent updates of INTEGRAL/IBIS and Swift/BAT surveys. 
The set consists of 9 objects, of which 6 were already known, two  are discussed here for the first time and one is proposed as a candidate GRG. 
In the new sample all, but one source, display an FR II radio morphology; the only exception is B21144+35B which is an FR I.
The objects belong to both type 1 and 2 AGN optical classes and have  redshifts 
in the range 0.06-0.35,  while  the radio sizes span from 0.7 to 1 Mpc.

Most of the sources in the sample have already been discussed in the literature except for 3 objects 
(SWIFT J1153.9+5848/RGB J1153+585, IGR J13107-5626/PMN J1310-5627 and SWIFT J0225.8+5946/WB J0226+5927) which are analysed in more detail here for the first time.

For  SWIFT J1153.9+5848/RGB J1153+585 we analyse available radio data, evaluate the flux density of each radio component  and further provide  observational evidences which suggest the presence of a blazar like core embedded in the large radio  structure.

IGR J13107-5626/PMN J1310-5627 has been optically classified as a type 2/absorbed  AGN for the first time in this works; its redshift is
0.093, making this a new GRG (size 0.7 Mpc). We also analyse the SUMSS radio image of the source, estimate the radio flux density of each component, evaluate its core dominance and indicate a possible flat core spectrum.

Finally for SWIFT J0225.8+5946/WB J0226+5927 we find  that it is a type 2/absorbed AGN of still unknown redshift.
In addition we analyse available radio data  and  estimate the radio flux density of each component.  The source core dominance is 0.15 and the core spectrum is likely flat.

Also for this set of objects, the  X-ray luminosity correlates with the radio core luminosity, as already found by
\citet{2018MNRAS.481.4250U} for our old sample of GRGs; the  correlation follow the 'efficient' branch of the radio -- X-ray correlation diagram, implying that the oft gamma-ray selection favours 
the detection of radiatively efficient objects.

Furthermore we confirm that in soft gamma-ray selected GRGs, the radio luminosity of the lobes is relatively low 
compared with the nuclear luminosity: this can be explained by a significant dimming of the radio lobes due to expansion
losses and/or by restarting activity, i.e. the sources are currently highly accreting and in a high luminosity state compared with the past activity 
that produced the old and extended radio lobes \citep{2018MNRAS.481.4250U}.

The only exception to this picture  is  PKS 0634-20, which is probably heavily absorbed in X-rays; if a  proper correction is applied to the core X-ray luminosity, then  the source behave like other GRGs in the sample.

As already done for our previous set of soft gamma-ray selected GRGs, we intend to follow up each of the source in the current sample in order to obtain spectral information particularly of the radio core and to study the radio morphology in search of signs of restarting activity.

\begin{figure}
\centering
SWIFT J1153.9+5848 \smallskip
\includegraphics[width=\columnwidth,trim={0 0.2cm 0 0.2cm},clip]{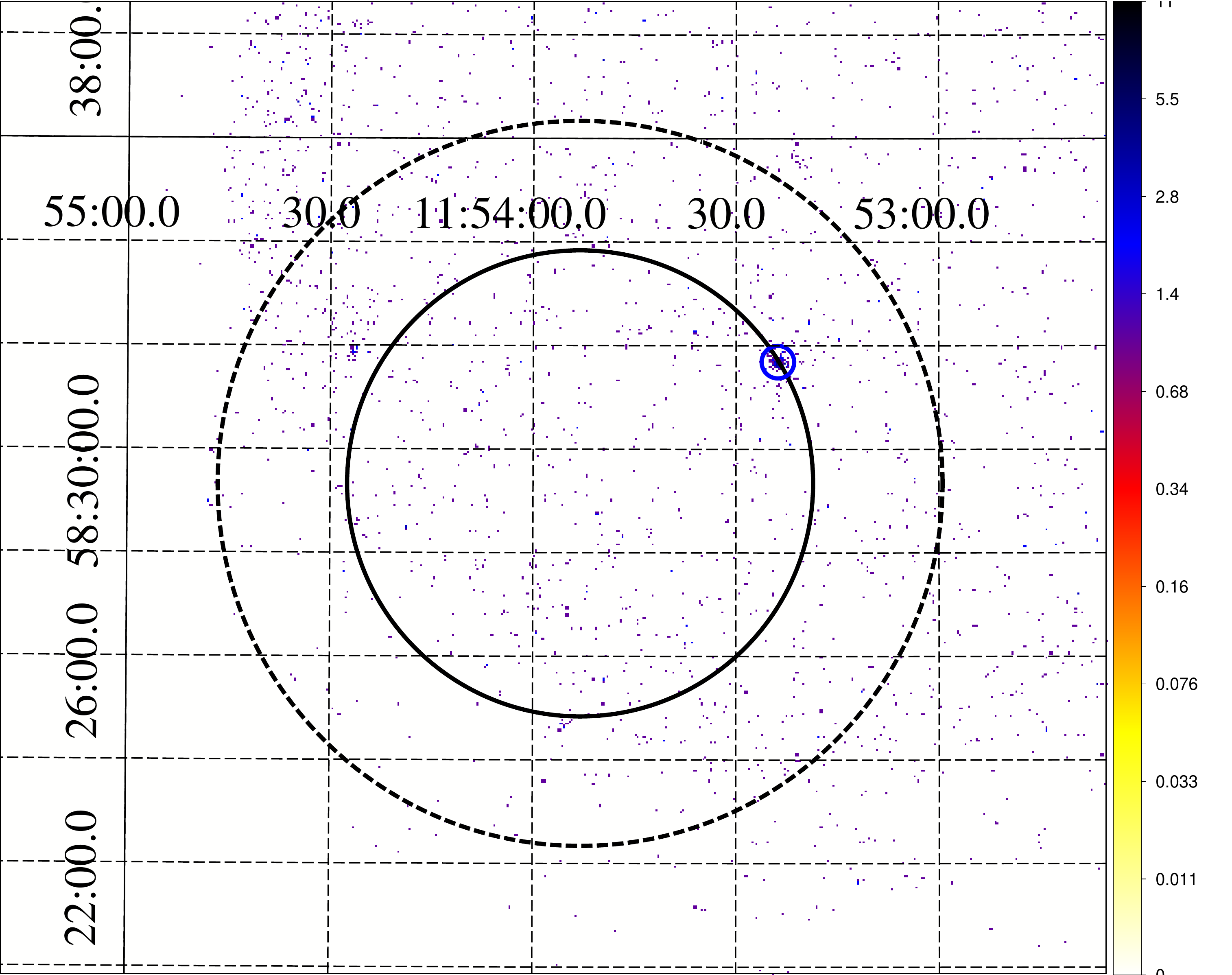}
 \caption{XRT image of the SWIFT J1153.9+5848 field, with BAT 90 and 99\% error circles.}
    \label{xrt1} 
\end{figure}

\begin{figure}
\centering
SWIFT J1503.7+6850 \smallskip
\includegraphics[width=\columnwidth,trim={0 0.2cm 0 0.2cm},clip]{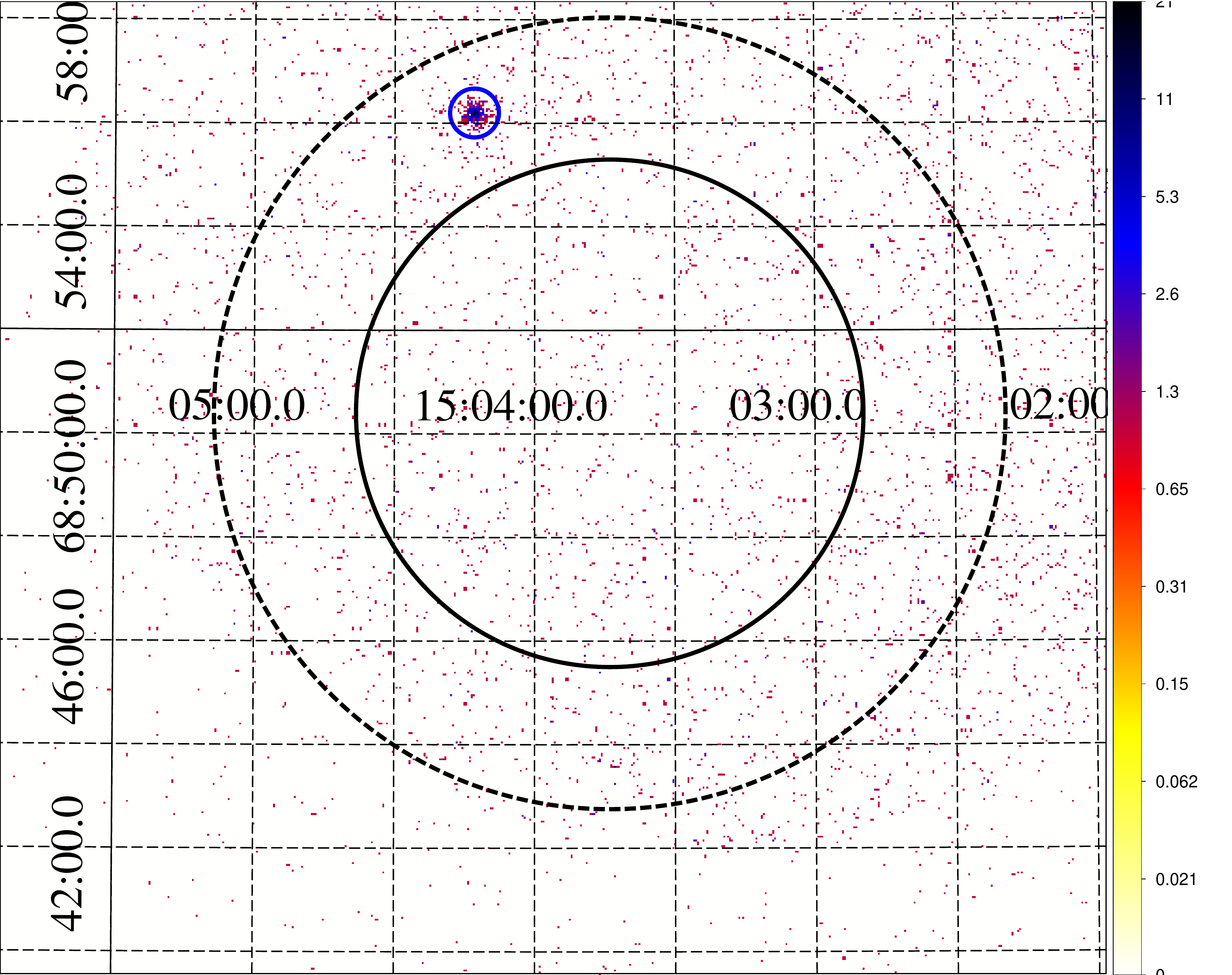}
 \caption{XRT image of the SWIFT J1503.7+6850 field, with BAT 90 and 99\% error circles.}
    \label{xrt2} 
\end{figure}

\appendix
\section{The X-ray data analysis}

XRT data, available for  8 GRG,  were reduced by means of the XRTDAS standard data pipeline package ({\sc xrtpipeline} v. 0.13.2) to produce screened event files. All data were extracted only in the 
Photon Counting (PC) mode, since this is the only mode that ensures a source fine positioning. 

We  stacked together all the available XRT pointings with exposure above 1 ks in order to enhance the signal-to-noise ratio and thus increase source detection probability. 

As a following step, we analysed the XRT images in the 0.3--10/3-10 keV energy bands by
means of the software package {\sc XIMAGE} v. 4.5.1 in order to substantiate  X-ray associations previously reported in the literature \citep{2018ApJS..235....4O, 2016MNRAS.460...19M, 2010MNRAS.403..945L}
and in the case of SWIFT J1503.7+6850 and SWIFT J1153.9+5848 to search for  the most likely soft X-ray counterpart.
We confirm all previous associations (X-ray positions and uncertainties as derived by our analysis are reported in Table~\ref{tab-app}1)
and also identify  the counterpart of the  two  still unidentified  sources.

In each of these two last cases,  we find a unique hard X-ray counterpart within the BAT 99$\%$ error box (see figure~\ref{xrt1} and  ~\ref{xrt2}) and report its  coordinates and  related errors in Table~\ref{tab-app}1; the X-ray positional uncertainty is  sufficiently small (few arcsec)  to allow the optical association of each source with an active galaxy,  RGB J1153+585 for SWIFT J1153.9+5848. and 4C+69.18  for SWIFT J1503.7+6850. \\
Next, we analysed the spectra of those GRGs for which the quality of the XRT data was good enough to perform a reliable spectral analysis, alternatively only an X-ray  flux value has been provided assuming a fixed photon index.
Source events were extracted within circular regions, centred on the source position, with a radius in the range 6--20 pixels (1 pixel $\sim$2.36 arcsec), chosen depending on the source brightness. 
Background events were extracted from a source-free region close to the X-ray source of interest. The spectra were obtained from the corresponding
event files using the {\sc XSELECT} v. 2.4c software and binned using {\sc grppha}
to have at least 1 count per bin, so that the Cash statistic could be applied.
We used version v.014 of the response matrices and
created individual ancillary response files \emph{arf} using {\sc xrtmkarf} v.0.6.3.\\
The only source having no XRT coverage (SWIFT J1238.4+5349), was observed by Chandra \citep[see also][]{2008ApJ...687..869S}. 
The Chandra ACIS data were reduced using the package \textsc{ciao} v4.11 and reprocessed with the standard tool \textsc{chandra\_repro}. 
Unfortunately  the BAT 90$\%$ error circle is only partly covered by the ACIS observation but despite this  Chandra detects one  source  whose coordinates are reported in Table~\ref{tab-app}1.  
Although we cannot exclude the possibility that other X-ray sources are present within the BAT positional uncertainty, we note that the only source detected is bright enough (its spectral shape extrapolated to the soft 
gamma-ray band provide a flux similar to the one observed by BAT) to be considered as a highly  probable association.


The source spectrum was extracted from a circular region with a radius of 2 arcsec, while the background was extracted from a circular region with a radius of 20 arcsec. The spectrum was then regrouped with the tool \textsc{dmgroup} to ensure at least 15 counts per bin.

For data fitting of all sources, we adopt a basic model consisting of 
a simple power law passing through Galactic absorption, unless intrinsic absorption was required as in the case of type 2 AGN.
In Table~\ref{tab-app}1, for each GRG observed  either by XRT or Chandra,  we report total exposure time on source and number of observations considered, the coordinates and relative uncertainties (at 90\% c.l.) of the corresponding X-ray counterpart, the count rate in the 0.3--10 keV energy range with relative sigma in parenthesis,  and the parameters of the best fit model (photon index and 2-10/0.2-12 keV flux).
Comparing with fluxes reported in the XMM-Slew survey (clean catalogue V2.0, \citealt{2008yCat..34800611S}), we note that our XRT fluxes are fully compatible with the range of values reported by XMM for a number of our sources; however the same comparison also indicates  some flux variability (a factor of 2-3) on yearly timescale in at least 4 objects (SWIFT J0632.1-5404, SWIFT J0801.7+4764, SWIFT J1238.4+5349   and SWIFT J1503.7+6850).

\begin{table*}
\footnotesize
\begin{threeparttable}
\caption{X-ray data analysis of new soft $\gamma$-ray selected GRGs}
\centering
\begin{tabular}{l l c c c  c c}
\hline\hline
Soft $\gamma$-ray Name &  Instr     &   Exp (Obs N)   & X-ray Position (error)        & Count rate (sigma)  & $\Gamma$  & F$_{2-10keV}$/  F$_{0.2-12keV}$ \\
                       &            &     Ks          & RA(J2000),Dec(J2000),(arcs)   & (0.3-10 keV) C/s    &           & 10$^{-13}$ erg cm$^{-2}$ s$^{-1}$\\
\hline\hline
SWIFT J0632.1-5404    & XRT         &  1.6 (1)        & 06 32 00.95,-54 04 55.89 (3.8) & 152.3$\pm$9.7 (15.7) & 1.92$\pm$ 0.15     & 40$\pm$10   /78$\pm$10 \\ 
SWIFT J0636.5-2036    & XRT         & 22.5 (7)        & 06 36 32.16,-20 34 53.35 (4.1) &   2.2$\pm$0.3 (7.3)   & 1.70$\pm$ 0.4      & 1.2$\pm$0.4 /1.8$\pm$0.6\\
SWIFT J0801.7+4764    & XRT         &  7.7 (1)        & 08 01 31.96,+47 36 15.43 (3.6) & 145.9$\pm$4.6 (31.7)  & 2.05$\pm$ 0.08     & 30$\pm$3    /65$\pm$4 \\
B2 1144+35B           & XRT         &  4.7 (1)        & 11 47 22.52,+35 01 08.55 (3.7) &  82.2$\pm$4.2 (19.6)   & 1.58$\pm$ 0.12     & 33$\pm$5    /55$\pm$7 \\
SWIFT J1153.9+5848(U3)& XRT         &  5.9 (4)        & 11 53 23.91,+58 31 40.65 (3.9) &  23.0$\pm$2.0 (11.5)   & 1.60$\pm$ 0.2       & 8$\pm$2     /14$\pm$3  \\
SWIFT J1238.4+5349    & Chandra     &  4.9 (1)        & 12 38 07.77,+53 25 55.99 (1.0)   & 284.0$\pm$8.0 (35.5)  & 1.79$\pm$ 0.09     & 13$\pm$2    /24$\pm$2\\    
SWIFT J1503.7+6850(U1)& XRT         &   4.1 (3)        & 15 04 13.11,+68 56 11.63 (3.7)  & 100.0$\pm$5.0 (20.0)  & 1.63$\pm$ 0.2     & 28$\pm$5    /51$\pm$6 \\
\hline
IGR J13107-5626      & XRT           &  6.6 (1)        & 13 10 36.90,-56 26 56.31 (4.9) &   5.5$\pm$0.9 (6.1) & 1.8f          & 14$\pm$5    /21$\pm$7   \\
\hline
SWIFT J0225.8+5946 & XRT       &  1.5 (1)        & 02 26 26.03,+59 27 48.15 (5.4) &  16.0$\pm$3.0 (5.3)   -   & 1.8f          &2$\pm$1/3$\pm$2      \\
\hline
\end{tabular}
\begin{tablenotes}
\item In IGR J13107--5626 an intrinsic column density of (3$^{+3}_{-1}$) $\times$ 10$^{23}$ at cm$^{-2}$ is present; 
\item In SWIFT J0225.8+5946 an intrinsic column density (3$^{+3}_{-2}$) $\times$ $10^{22}$ at cm$^{-2}$ 
is present.
\end{tablenotes}
\end{threeparttable}
\label{tab-app}
\end{table*}

\section{The optical data analysis}
We retrieved spectroscopic observations of the optical counterpart of IGR J13107-5626 (see \citealt{2010MNRAS.403..945L} and Appendix A  for the source optical/X-ray coordinates)
through the ESO archive\footnote{\url{http://archive.eso.org/eso/eso_archive_main.html}}.  IGR J13107-5626 was observed on 2019 June 19 with the EFOSC2 instrument \citep{1984Msngr..38....9B}, equipped with a 
Loral/Lesser 2048$\times$2048 pixels CCD, mounted on the 3.58 m 
New Technology Telescope of the ESO-La Silla Observatory (Chile). 
Three 1200s exposures were acquired in 2$\times$2 binning mode
starting at 01:48 UT with the grism \#13 and a 1$''$ slit: this 
setup afforded a dispersion of 5.5 \AA/pixel.

The observations were bias corrected, flat-fielded, cleaned for 
cosmic rays, background-subtracted and the corresponding spectra
were extracted following standard procedures \citep{1986PASP...98..609H} using 
IRAF\footnote{IRAF is the Image Reduction and Analysis Facility, 
distributed by the National Optical Astronomy Observatories, which 
are operated by the Association of Universities for Research in 
Astronomy, Inc., under cooperative agreement with the National 
Science Foundation. It is available at {\tt http://iraf.noao.edu.}}. 

All calibration frames were acquired on the day preceding the 
observing night. Wavelength and flux calibrations were performed 
with the use of lamp acquisitions taken before the science data 
frames, and of the spectrophotometric standard star LTT 7379
\citep{1992PASP..104..533H,1994PASP..106..566H}, respectively. The uncertainty in the 
wavelength calibration was of 3 \AA. All spectra were eventually 
stacked together to increase the final signal-to-noise ratio.

\section*{Acknowledgements}
 The authors acknowledge financial support from ASI under contract n. 2019-35-HH.0., in particular for  G.B. research contract. We thank the anonymous referee for insightful comments and suggestions that improved the quality of the paper.
 
 \section*{Data availability}
 The data underlying this article are available in the article. Data not specifically appearing in the  article like the products of the X-ray analysis (spectra and images) will be shared on reasonable request to the corresponding author.
 
\bibliographystyle{mnras}
\bibliography{grg-biblio} 

\end{document}